# Research Paper - "Just a little bit on the outside for the whole time": Social belonging confidence and the persistence of Machine Learning and Artificial Intelligence students

Katherine Mao[1], Sharon Ferguson[1], James Magarian[2], Alison Olechowski[1]
[1] University of Toronto
[2] Massachusetts Institute of Technology

**Abstract**
The growing field of machine learning (ML) and artificial intelligence (AI) presents a unique and unexplored case within persistence research, meaning it is unclear how past findings from engineering will apply to this developing field. We conduct an exploratory study to gain an initial understanding of persistence in this field and identify fruitful directions for future work. One factor that has been shown to predict persistence in engineering is belonging; we study belonging through the lens of confidence, and discuss how attention to social belonging confidence may help to increase diversity in the profession. In this research paper, we conduct a small set of interviews with students in ML/AI courses. Thematic analysis of these interviews revealed initial differences in how students see a career in ML/AI, which diverge based on interest and programming confidence. We identified how exposure and initiation, the interpretation of ML and AI field boundaries, and beliefs of the skills required to succeed may influence students' intentions to persist. We discuss differences in how students describe being motivated by social belonging and the importance of close mentorship. We motivate further persistence research in ML/AI with particular focus on social belonging and close mentorship, the role of intersectional identity, and introductory ML/AI courses.

**Introduction**
Artificial intelligence (AI) is now used in almost every industry [1]. As such, ML/AI courses, majors and careers are increasingly sought out by university graduates. While ML/AI falls within the boundaries of Science, Technology, Engineering and Math (STEM), there are elements of this new field, industry and type of work which set it apart; Thus, a study dedicated to understanding the dynamics of student perceptions of ML/AI can help us better evaluate how the field may be encouraging or discouraging broad participation. This insight would ultimately inform our understanding of student persistence, where persistence refers to one's continued membership or association, through academic study or occupational work, with that particular profession.

Social belonging, or a sense of belonging in the social and cultural aspects of a profession, has been shown to predict student persistence [2]–[5]. Studies find that women students in STEM report lower levels of social belonging compared to men [2]–[5] and that belonging confidence is a stronger predictor of persistence for women than men [3]. In engineering contexts, this disparity in belonging stems from differences in *professional socialization*, the process of acquiring the knowledge and skills of a profession, aligning with the profession's values, and developing a professional identity [6]. Thus, the study of social belonging confidence in students, particularly through professional socialization, is a key step in understanding persistence in the field of ML/AI. We study social belonging through the lens of confidence, another predictor of persistence [7]–[9]. We define social belonging confidence as how confident a person feels that they will fit in with the social and cultural aspects of a profession and develop meaningful relationships with their peers. *Belonging uncertainty*, defined by Walton and Cohen [10], may be interpreted as a lack of this confidence.

Lower levels of confidence have been found to negatively affect persistence of women and girls in highly male-dominant fields like engineering and computer science (CS) [7]–[9]. Recent work suggests that the importance of *technical confidence* on persistence seems to be diminishing [12], [13], and other confidences, like *professional role confidence,* play a larger role in persistence [7], [11]. In ML/AI, professional role confidence was found to be a strong positive predictor of intentional persistence [14]. Social belonging confidence, however, remains unexplored.

There are unique aspects of ML/AI as a field that point to the need for such a dedicated study. ML/AI has an increased emphasis on non-technical skills [15], [16], ambiguous ethical implications [17], [18], [19, p. 50], and can be highly competitive and lucrative [20]–[22]. We conjecture that these features may result in student persistence being driven in different ways than in traditional engineering. Further, most studies of persistence look at drop-out rates [3], [5], [7], [9] where persistence is the expectation. Since there are currently few degree programs in ML/AI, there are a plethora of ways a student may end up working in ML/AI, and we know little about these pathways.

Statistics show that the field of ML/AI lacks diversity; currently only 32% of AI professionals are women [23], and though field-specific assessments of racial demographics are still underway, only 7% of professionals in broader computer occupations are Black [24]. This lack of diversity may have direct impacts on the outputs of the ML/AI field, where training data and algorithms are often encoded with the societal and cognitive biases of their creators [25]. If left unmanaged, these systems can perpetuate stereotypes and discriminatory practices coded in existing data [25]. Increasing diversity in the field of ML/AI is crucial in managing these biases [20], [25], and diversifying the field requires improving learners' sense of belonging. Considering the impact AI has in our world, we must study the reasons for low participation rates of women and racial minorities in ML/AI.

Here we present an exploratory research study of student persistence in ML/AI, an understudied field in persistence research, to identify important directions for future work. We investigate how social belonging confidence is associated with intentional persistence of students of ML/AI courses, and how these students develop this confidence. Via a small set of semi-structured interviews and a thematic analysis, we examine high and low persistence intentions among students in ML/AI classes to assess the role of social belonging confidence toward their career aspirations. Our findings identify areas of future research to gain critical insight into what the field of ML/AI can do to build social belonging confidence, thus attracting and retaining people of diverse identities to the profession.

## Background

### Persistence
Cech et al. [7] describe two components of persistence: behavioural persistence, defined as the achievement of the requisite credentials (a degree or job, for example); and intentional persistence, defined as a commitment to work in a profession in the near future. Factors affecting persistence have been studied to understand the "leaky pipeline" phenomenon of underrepresented groups leaving STEM fields at the pre-university level [26], [27], the early undergraduate level [5], [9], [28], the senior undergraduate or graduate level [3], [14], and the working years [11]. It is important to study persistence at different stages. Studies of the early years indicate that low technical confidence in women is a strong negative predictor of persistence in STEM [9], while those focused on later years indicate that social belonging has a stronger relationship [3], [11]. Since ML/AI is a specialized field, students may enter university under a general program, and only study ML/AI at the senior undergraduate or graduate level. Our work will therefore focus on these later years.

**Social Belonging Confidence**
Confidence is a broad term referring to how certain a person feels about their abilities or choices. Differences in self-confidence between men and women have been cited as a major factor in differences in persistence in STEM [7]–[9]. Existing studies have looked at confidence in one's skills required in a field (*technical confidence*), [7]–[9], [11]–[13], [29], [30] and confidence that a chosen career is the right path for an individual (*professional role confidence*) [7], [14].

Technical confidence rests upon beliefs in one's skills or abilities pertinent to a field. Historically, a major deterrent for women considering STEM has been lower levels of math self-assessment [8], though recent studies suggest that this is becoming less important for persistence [12], [13]. For instance, researchers found that students' have broadened their perception of the skills required in CS [13] with more awareness of the non-technical or "soft" skills, such as reliability and teamwork, that are sought by employers [29]. This shift suggests a need to explore how students perceive the field of ML/AI and highlights the importance of studying other factors contributing to persistence.

Professional role confidence, comprised of *expertise confidence* and *career-fit confidence,* is the degree to which a person feels confident in their competence in, satisfaction with, and identity within a profession [7]. Expertise confidence is a holistic view of one's confidence in the skills and knowledge required to succeed in a profession [7]. Career-fit confidence is the degree to which the day-to-day work and values of a profession align with the interests and beliefs of an individual [7]. Both dimensions are strong predictors of persistence in students of STEM [7] and ML/AI [14]. Seron *et al.* argue that these confidences are shaped by *professional socialization* [6], [7]. This study will continue the inquiry into how confidence is developed through professional socialization in ML/AI.

Another contributor to persistence is belonging [31], [32]. In STEM fields, a sense of belonging has been shown to be a positive predictor for career interest [26], [27] and belonging uncertainty is a negative predictor for intentional persistence [3], [4], [28]. Belonging uncertainty refers to situations where "members of socially stigmatized groups are more uncertain of the quality of their social bonds and thus more sensitive to issues of social belonging" [10, p. 82]. As it is closely tied to stereotype threat and discrimination, marginalized groups are at higher risk of experiencing belonging uncertainty [33]. Women may also feel alienated in the highly "masculine"[3] cultures of STEM. Computer science and engineering are perceived to be the most masculine and have the poorest gender parity among STEM fields [34]. Some expectations in these environments may include overworking, social awkwardness, or being unkempt [35], which may alienate women because they are less aligned with how they view themselves or are expected to behave [34]. Often, those who do persist learn to adopt these traits to divert unwanted attention, which may distance them from other women [35] and further normalize these cultures. To improve social belonging, interventions suggested include fostering a strong STEM identity [36], external validation of their competency [37], inclusive department events in university [4], and relatable role models [34].

---

[3] The term masculine is in quotations because the described traits and cultures may not necessarily be intrinsically masculine. Rather, the described traits are socially coded to be masculine or non-feminine.

Prior findings of social belonging's salience as a predictor of persistence in STEM fields suggests that it deserves a thorough investigation in ML/AI. In this work, we begin the investigation of social belonging through the lens of confidence. This study examines the following research questions:
- *RQ1: What motivates students to intend to persist in the field of ML/AI?*
- *RQ2: How does social belonging confidence affect ML/AI students' intentional persistence?*
- *RQ3: What kinds of experiences develop social belonging confidence in students of ML/AI?*

**Methods**

This qualitative study builds upon a related survey study [38]. The survey (n=165) was distributed to students during ML/AI classes at a major Canadian university. This survey collected information regarding students' intentional persistence in ML/AI, their confidence in various skills and social belonging, their interest in work with a positive social benefit, their motivations for taking ML/AI courses, and demographic characteristics. Participants for the current interview study were solicited on a volunteer basis from the survey respondents, which enabled us to describe our population in more detail using the survey responses. Of the 165 students surveyed, 75 students expressed interest in participating in the interview, with only 13 responding to a follow up demographic questionnaire, and 8 ultimately completing an interview. The relatively low secondary response rate may be due to the time gap between the survey and interview study. The gap in time, change in semesters, and ramping up of new courses may underlie the decreased interest. Both the survey and interview protocols were approved by the university's research ethics board.

**Participant Characteristics**

Participant characteristics (**Table 1**) were obtained from their demographic questionnaire and survey responses; all are self-reported. Race/ethnicity is measured by asking participants whether they identify as one or more of a set of visible minorities, a standard census question in Canada [39]. Women are over-represented in this subsample compared to the larger survey demographics (50% of our sample vs. 33% in [38]); Chinese and South Asian are the largest visible minorities in both (25% each vs. 42% and 18%, respectively); Mechanical & Industrial and Computer Engineering were the most common majors (62.5% and 12.5% vs. 42% and 25%, respectively); and graduate students were underrepresented in this subsample (25% vs 43%).

There were also non-demographic characteristics we can use to describe the sample, based on the survey responses. Six participants were pursuing a major, minor, certificate or specialization in ML/AI; they reported a high level of short-term intentional persistence on the survey, indicating that they intended to take another ML/AI course. This sample ranked themselves very high on *social benefit interest*, a measure of how important it is for one's work to positively benefit society [40]. Further, these participants had higher non-technical self-assessment (communication, teamwork, and leadership [38]), expertise confidence [7], competitive participation (how often students participated in competitive events or activities [38]), and ratings of importance of earning a high salary. On the other hand, the interview participants had lower levels of career-fit confidence [7], and social belonging confidence. Further, they rated their environment as less toxic (discrimination and unequal treatment [38]), and they were less likely to be motivated to take an ML/AI course based on the popularity of the subject. The sample had independent variable averages that were within the standard deviation of the larger sample (see **Appendix A** for mean participant responses, and **Appendix B** for the survey measures). To best understand the factors that impact the decision to remain in or leave the field, we recruited both highly persistent and non-persistent participants.

**Interview Procedure**
The interviews were semi-structured, one hour in duration, and conducted online over Zoom. The interview opened with questions about the participant's *interests and intentions to persist in ML/AI*. These questions provided an initial sense of their commitment to this field. Then, the interview focused on the interviewee's experiences in *professional socialization*, such as coursework, student teams, and related work experience. These questions aimed to understand the participant's experience in ML/AI, particularly through interactions with other people, and how those experiences contributed to feelings of belonging. Then, we asked about *social belonging confidence:* how important a sense of belonging in their career is, how much they feel like they belong in ML/AI education, and how much they feel like they will belong in the field of ML/AI, should they stay.

**Table 1**: Participant Characteristics. # = number of participants (out of eight)

| Factor | # | Factor | # |
|---|---|---|---|
| **Gender*** | | **Program** | |
| Woman | 4 | Undergraduate | |
| Man | 4 | Industrial Engineering | 4 |
| Non-binary, genderfluid or Two-Spirit | 1 | Computer Engineering | 1 |
| **Visible Minority** | | Engineering Science | 1 |
| Chinese | 2 | Masters of Engineering | |
| South Asian | 2 | Aerospace Engineering | 1 |
| Other | 1 | Mechanical Engineering | 1 |
| Latin American | 1 | **Year** | |
| Not a Visible Minority | 2 | Undergrad – Year 3 | 3 |
| | | Undergrad – Year 4 | 3 |
| | | Graduate | 2 |

* One participant identified as more than one gender.

**Thematic Analysis**
Interview transcripts were uploaded into the NVivo 12 software for analysis. The first author completed the analysis, following the Braun and Clarke [41] method. After reading over the transcripts to understand the data, the next stage was open coding, which involved labeling quotes with initial codes, capturing the topic and associated feelings. For example, this quote was coded with "programming", "writing," "prefer programming", and "fun":

> I enjoy having experience on both sides, although I definitely prefer doing the code as it's…just easier. The communication's always hard…it's just something I've had to do so much of, and I really like the code as it's always fun and unique...

The initial coding phase revealed about 300 open codes, which were then grouped and organized hierarchically in the NVivo platform, resulting in 13 high-level codes (**Appendix C**). The quotes within each high-level code were reviewed to generate themes in relation to our research questions; these themes represent a complete thought that puts the open codes in a broader context. Previous interviews were revisited when new codes emerged. Due to the semi-structured nature of the interview, minor modifications (i.e., follow-up questions) were added to the interview guide in response to emergent topics. Five themes, that relate to our research questions, were generated: *Exposure and initiation, perception of careers in ML/AI, Social belonging and motivation, importance of mentorship, and perceptions of changing cultures*. The mapping of high-level codes to themes is shown in **Appendix C**. The first author and the last author discussed the codes and themes, solidifying their definitions ("Reviewing Themes", or step 4 in the [41]). We organized themes around the research questions. Excerpts are presented with minor edits for clarity and privacy.

## Results

**RQ1: What Motivates Students to Intend to Persist in ML/AI?**
*Exposure and Initiation*: Our first theme relates to participants' exposure to ML/AI and their initial motivations to enroll in ML/AI courses. As these students were pursuing technically oriented degrees, some exposure came through the academic environment—it was often their first hands-on encounter with the material, but they had heard of the field prior. Most participants cited they were motivated in part due to the popularity and employability of ML/AI skills. Some are still exploring their career aspirations, meaning that they took ML/AI courses with low commitment to the field. Further, some participants had heard about ML/AI, but did not see it as a possible career option until they were encouraged to try it during their undergraduate studies. For example, this participant was exposed to ML/AI in high school, but thought it was "out of [their] scope" until university, "before university, I was like very not a programmer…I thought "Oh, I'm not smart enough to do that"…One of my friends…was showing me this awesome Shakespeare script generator…I always knew that this was a really cool thing…it just seemed like something that I couldn't do…"

Others encountered ML/AI outside of the classroom, such as through projects in non-ML/AI internships, or extra-curricular design teams. Regardless of how they were first introduced, all participants sought to take an introductory ML/AI course in university. Some suggested that the culture in introductory courses makes it difficult to engage with the material. For instance, this woman-identifying participant shared that the male-dominated nature of one ML/AI course contributed to feelings of imposter syndrome that prevented her from asking questions in class:

> Gender stereotypes contribute oftentimes to the imposter syndrome I feel. … Something about the [ML/AI] course feels very male-dominated, like I often find during lecture, it is like a lot of just male students asking really difficult questions…I think that makes me feel kind of like scared of the prof, because I don't want to ask him really basic questions…

Most participants agreed that while courses discussed ML/AI research, they covered very little about ML/AI in industry, "I think it would help more if I had the chance to like intern in the ML field in industry, because I think there's still some gaps that exist between like schoolwork and work in industry." Additionally, many participants mentioned using online resources (such as courses, blogs, articles, newsletters, and question boards) to complete ML/AI assignments and develop skills.

*Perceptions of careers in ML/AI:* Next, we examine students' perceptions about careers in ML/AI. While we bundle the terms together, as has been done before [42], some participants made a distinction between ML and AI. Our participants described ML as more technical, with an emphasis on programming skills, and involving the development of algorithms. AI, on the other hand, appears to be used as a more general term which encompasses data science, robotics, and applications of ML in finance and healthcare. Notably, those participants who are more passionate about ML are the most confident about their persistence in the field. These individuals described enjoying programming and having a deep appreciation for the technical aspects of ML/AI. On the contrary, the participants who shared interest in applications of AI expressed human-focused motivations, and are interested in using ML/AI to solve complicated problems, but they do not wish to be designing models, as they find less enjoyment in programming. Participants in this category report moderate to low persistence. This participant, who was initially drawn in by the applications of AI, found that ML course content didn't quite match up with their expectations:

> I feel like I just don't see the practical applications of [ML/AI] as much, where it kind of feels like we're just putting random code into the computer and then like outputting random solutions…I think maybe there's like a bit less of a human side, which is what I had originally been interested in...I guess I feel like it's a lot of buzzwords that are not that interesting to me…
> I like strategy and analytics, I don't think I would go any further than that into machine learning.

When asked what students would pursue if not ML/AI, participants suggested consulting, data modelling, product management, robotics, and software engineering. There was also a common sentiment that these "alternative" pathways are not necessarily at odds with a career in ML/AI. For some, these pathways are what they are working towards anyway; their decision to "persist in ML/AI" is more a question of whether the specific role they seek includes an ML/AI component.

Participants shared that both technical and non-technical skills were important for a person to be successful in ML/AI. First, all participants identified programming as an obvious technical requirement, as well as math and statistics fundamentals. In terms of non-technical skills, communication, particularly to a non-technical audience, was emphasised, "being able to communicate is really important...because of how difficult it can be to get people to understand what [AI] really is…many people just view it as a magic box that solves their problems and [it's] so important to communicate that that's not what it is."

Students also discussed the skill of understanding how technology fits into the bigger picture. This was described as being "socially conscious," "well rounded," and "knowing what problems to work on." Participants discussed being intentional about their roles and avoiding unethical AI applications:

> I would never want to be in a position where the tools I was using were being used in a discriminatory or inequitable way… I don't want to work with like identifying faces in an airport...
> I kind of like keeping it with the numbers…like high-level system operations within hospitals…
> I definitely want to use the tools in a way that you know like promotes equality and care for people.

Many participants were more confident in their non-technical skills than their technical skills. Some developed this technical confidence over time, through mandatory courses. For example, developing programming confidence was critical for this participant:

> I think I just got the hang of it…I was like "Oh, maybe I can do this" because it seems my skills are up to par now…it was more like I knew that my skills weren't there [initially] to be able to pursue AI/ML. But then afterwards once I had those skills because I was forced to [develop them] from the engineering curriculum, I realized that I could do it...

However, for some, perseverance was not enough to develop these skills. The following individual recognized their perseverance and growth mindset, but felt it wasn't enough:

> I felt like I've been around people who I feel like have this better natural ability than I do, but I'm willing to like to put in the time and effort to try to do well. But I think I've also reached a point in my life where I'm kind of like tired of feeling like I have to catch up to other people and I'd rather spend my energy on doing things that come easier to me and make me happier… Even though I like the industry, [and] I do like coding, it's just this fact that I feel so slow in it...

**RQ2: How Does Social Belonging Confidence Affect ML/AI Students' Intentional Persistence?**
*Social Belonging and Motivation*: When asked to describe an ideal workplace, most participants did not describe social connections and instead focused on areas such as organizational structure and work-life balance. However, when prompted about social belonging, all of them stated that it was something they valued. Participants with different needs prioritize different kinds of social connection: participants that self-identified as extroverts described placing a greater emphasis on finding community, casual conversations, and forming friendships; those participants who described themselves as introverts valued fitting in, open communication, and recognition of their work. For the self-described extroverts in our sample, having a sense of belonging may not be a requirement in their future workplace, but it may be a motivator to go above and beyond: *"I can work, even if I don't feel like I belong because like, that's like my commitment… I will do my work, but…to go above and beyond in the work, I feel like a sense of belonging is necessary..."*

Similarly, one participant explained that social connections helped them overcome technical challenges by being able to work through problems with their peers:

> When I was doing like my coding assignments in second year, and like isolated at home, that was very disheartening...But then third year has been my first time back in person and I love it…And I think that [the community] really helps me get through every aspect of engineering…also the technical components of coding and stuff because I just like being around my friends and be able to talk through problems and even just like, recognizing that other people actually struggle as well.

Participants in our sample who described themselves as introverted also discussed valuing social belonging; though it seemed to not impact their work as much. The first example is from a participant who completed an internship at a company where they lacked connection:

> I feel like [having people I could connect with] would have brought a level of comfort. So as good as the internship was, there's always…a little bit of awkwardness… a certain culture there that I wasn't really a part of…I was just a little bit on the outside for the whole time… I don't know if it would have improved my work…but it just would have made the experience a lot more pleasant.

Other participants elaborated on the impact of social engagement. In classes, one participant described that forming a relationship with an instructor would increase feelings of belonging in the research community, but would not affect their performance in the course. Another participant shared how they would rather be judged on merit at work, as opposed to their fit with others.

**RQ3: What Kinds of Experiences Develop Social Belonging Confidence in Students of ML/AI?**
*Importance of Social Connection through Close Mentorship:* Social connection through close mentorship appears to be highly salient toward developing social belonging. Almost every participant interviewed was able to recall a person who has strengthened their sense of belonging and confidence in ML/AI. The first example is from a participant who intended to persist only in analytics and strategy roles. Having an upper year mentor from an extracurricular group, who demonstrated the value of non-technical skills changed their view of their future career:

> The greatest strengths I ever saw from her was her leadership abilities and communication skills...it was nice to see that those skills could be rewarded in her career…this shocking revelation I had once was when she told me her GPA, and…it wasn't that great. But she was someone who I thought was incredibly successful and talented, and it just made me realize… maybe like skills other than [programming] could be important in my career.

This social connection helped break down the feelings of uncertainty tied to the participant's low technical confidence. By demonstrating alternative strengths that were valuable to the participant's field of interest, this mentor figure broke down some of the participant's prior beliefs of how candidates are valued in careers. Another example shows how social connection through relatable mentors can also be directly impactful to one's career. One participant described her experience at an ML/AI consulting internship. The work environment was built on networking to get on projects, and she found that the coworkers who saw themselves in her were the most willing to help:

> All of the people who really liked me were other brown women…They were a lot more eager to help, like at the end of the conversation will be like … "Definitely reach out if you want any work, I could… get you plugged into whatever [project]". She…helped me get on the project, which at [company] is important, because the only way you get onto your project is by knowing people

While this participant doesn't believe that her experiences from this company are representative of ML/AI as a whole nor that these were the only people who she formed a connection with during her internship, it reveals that relating to others can deepen the connections with them, intentionally or otherwise. In this case, it provided this participant with career-advancing opportunities. For others, role models provide the confidence to pursue ML/AI, even from an unconventional background. This participant said that they had hesitations about pursuing ML/AI coming from mechanical engineering, but interactions with some professors have strengthened their belonging confidence, "being a mechanical engineering professor but focusing on robotics...seeing that was a good role model for me…it doesn't matter if I'm mechanical, I can still focus on software and robotics."

*Perceptions of engrained and changing cultures within the discipline:* The last key theme identified was the sentiment among the participants that experiencing alienation in highly masculine cultures is broadly expected across technology-intensive industries, and has less influence on persistence compared to other factors:

> [There are] spaces that are very men dominant and potentially not super welcoming to women of color, but you don't just get that in tech; you get that in almost every industry…in terms of whether I think I belong in ML/AI, not any more than I would belong in finance or some other field. But I think you can find places you belong if you like are intentional

Another participant, who self-identifies as queer and is highly persistent towards ML/AI, describes how a lack of diversity can contribute toward a 'bro culture' in engineering:

> What makes me feel a little uncomfortable with the 'bro culture' stuff …I am a queer engineer, and I've definitely had to experience a lot of that kind of masculinity before in engineering...I'm just not a big fan of it, and so like kind of having to sit there and just pretend to be like, "Yeah, I know what this is. I like this," it always feels very awkward...When I think about diversity, I think about avoiding these kinds of toxic cultures that can come up when you get a group of people who have a very unchallenged life experience and are very assumptive with that.

For these participants, social belonging is not as important for persistence compared to other factors like interest. There is a sense of acceptance of this male-dominated culture that comes with engineering, and so specializing in ML/AI does not prevent them from experiencing toxic environments stemming from a lack of diversity. Rather, they can be intentional about the specific company environments that they seek out to find places where they do belong. Still, the latter participant hopes to find community with people of similar technical interests:

> I think [community] is going to be really important for just like professional development … feeling like I belong where I work and having that sense of community… I doubt that I'm going to get the kind of queer engineer sphere around me just in a workplace…I'm hoping that I can get that from just working with people who have very similar technical interests.

Some participants, offering an alternate perspective, believe that the fast-growing nature of the field means that a set culture has not been formed yet, and it is continually being shaped, "[ML/AI] is growing at such a rapid pace. It's also like pretty new as an industry…So, you see a lot of like fresh faces and newcomers and I feel like the community is really open to new ideas. Like it's constantly improving. … it's not very stagnant."

In summary, the insights developed from this exploratory study are as follows:
- Initial introduction and encouragement in ML/AI seem to be important for persistence. Introductory programming courses are key prerequisites for upper-level ML/AI courses, but are sometimes described as intimidating.
- Our participants understand ML to be more technical, and AI to be more application. These perceptions influence what they consider persisting in ML/AI.
- Programming skills and perseverance are described as important for persisting in ML/AI; our participants used external resources and worked on side projects to build these skills.
- Self-described extroverts in our sample explained that social belonging in the workplace may be a motivator, whereas self-described introverts were motivated by interest.
- Relatable mentors helped our participants to develop social belonging in ML/AI.
- While our participants commented on the male-dominated culture in ML/AI, this did not seem to influence social belonging as they perceive this culture to span many STEM fields.

In the following section, we review each of these insights in turn and discuss how they relate to past work and inspire future research directions.

## **Discussion**

The insights presented here, derived from a small sample, are not generalized conclusions on their own, but highlight important areas for further study. We identified the theme of exposure and initiation to ML/AI courses, and how programming may influence students' intentions to persist. Our participants described self-sorting into ML/AI roles based on interest and comfort with programming. Programming skills seemed to be a barrier for many participants at first, but most described that having a growth mindset allowed them to overcome this barrier. Having a fixed mindset, or believing that intelligence and talent are innate, has been shown to be associated with feelings of imposer phenomenon [43], belonging uncertainty [44], and stereotype threat in women in STEM [45]. Women are also more likely to believe in "brilliance" or innate field-specific ability in male-dominated fields [44]. Furthermore, women are less likely to be exposed to programming in high school, making them feel behind their peers in introductory courses [34]. These factors may make women more vulnerable to developing a fixed mindset in ML/AI. If we consider this with what Seron *et al.* describe as the "hegemony of meritocratic ideology", the belief that success is necessarily a result of one's ability, having a fixed mindset may reinforce ideas about the myths of equal opportunity and necessary trade-offs between diversity and quality, which systemically bar women and other minorities from these fields [46]. These beliefs can be shaped by culture and professional socialization [6]. Thus, the question of how to introduce students to new fields is important in shaping mindsets. Some of our participants described improved confidence after introductory programming courses. However, for others, these courses seemed to reinforce negative beliefs. This is similar to the "weeding-out" effect of introductory courses [47], like calculus courses in engineering [9]. Although these ML/AI courses are not mandatory, nor is it the first time these students encounter programming, they perhaps still have a similar "weeding-out" effect.

With regards to student intentions to persist in ML/AI, we found that those more interested in highly technical ML/AI seem to be the most confident in their persistence. This may suggest that the field has well defined pathways for highly technical roles, but may not for less technical ones, leaving students unsure about what the pathway looks like. In subsequent quantitative studies, we should clarify specific roles in ML/AI, to reveal insights like gendered patterns in occupational segregation.

Our participants described often relying on online resources to learn ML/AI, perhaps indicating that academic presentation of the content is not enough to build confidence. Perhaps because of the fast-moving nature of the field and high skills bar of entry, the community creates resources to fill a gap in traditional education streams. Individuals may require a certain level of self-motivation to seek out these online resources. While it might seem appropriate for a highly competitive field to attract these traits, we must look at the nature of this self-motivation. Those who are more motivated by technical challenges may be excited to take on personal projects and tinker with ML/AI. However, those who are motivated to use technology to solve real-world problems may not have the opportunity to gain that experience or may not be interested in doing ML/AI side projects if the social impact is not realized. Prior work has revealed that women tend to value work with a sense of purpose [18], thus an overemphasis on technical projects and independent work can further alienate women. Further, students who work part-time or have other commitments at home may not have the time to work on these side projects, perhaps creating a socio-economic gap in this field.

Most participants described that social belonging confidence did not strongly influence their intention to persist, inconsistent with previous studies in engineering [3], [4], [28]. However, closer inspection revealed that attitudes surrounding social belonging were still important for motivation. Among our participants, self-described extroverts said their motivation was negatively impacted by a lack of social belonging, while the self-described introverts may gain a sense of belonging through their work instead. Thus, it's important to understand different ways people develop social belonging.

In this vein, all of our participants described that close mentorship built social belonging, making it the most impactful source of social belonging from our study. Most of the participants described connections with people of the same gender, race, or those with similar extracurricular interests and career aspirations. This demonstrates the impact of the "relatable role model". A lack of women role models has been shown to affect women's STEM aspirations [34], and the relatability of these role models adds to a person's sense of belonging and persistence [34]. Additionally, we heard that our participants expect masculine cultures as part of all fields within engineering. Considered along with the previous findings, this suggests that it may be more difficult for women and other minorities to develop close mentorship in a male-dominant field like ML/AI. Thus, the people who may be most likely to persist either fit the existing culture and can find a mentor, or are independently motivated and interested in the technical aspects of the field. This means that minorities, or those who require social connection to stay motivated, may be less likely to persist in ML/AI, as shown in previous work [38]. At the same time, some participants described the expectation of a fresh and fluid ML/AI culture, making this a particularly opportune time to increase diversity and inclusion efforts.

As a summative example of our insight, we highlight one participant, who, despite aligning more with the traits of the low persistence group in our sample, described themselves as persistent. They are extroverted, motivated by developing technology to help people, and less interested in highly technical roles. However, through close and consistent connection, they were able to develop programming confidence and an overall positive outlook on their career in ML/AI. This individual discussed technical roles not as obstacles, but as a stepping stones, and highlighted the role of their friends in overcoming challenges. They advocated for lowering barriers to programming:

> ML and AI are something that I've found a huge interest for, and I think are very important tools, and are going to help me, in wherever I go, but if I didn't have my buddies to kind of push me along, I never would have stumbled upon it…it's important to give people the time to figure out how to program…more people will have the opportunity to realize how cool these things can be.

## **Implications & Future Work**

Based on the early insights discussed, we highlight three considerations for educators as we continue to develop ML/AI educational programming in response to the growing demand: promoting meaningful social connection, defining and sharing numerous pathways within ML/AI, and thinking carefully about building programming confidence and avoiding the weed-out effect.

Prior work has shown that social belonging intervention through worksheet-style materials improved performance, confidence, health, and social engagement, particularly for women and Black students in engineering and CS [48], [49]. Our study proposes a potential alternative form of social belonging intervention through diverse opportunities for social connection. Extroverted individuals may enjoy weekly socials, but introverted individuals may prefer 1-on-1 mentorship. Institutions should create opportunities for individuals to find role models with whom they have things in common, including gender and beyond, since role models' relatability may play a role in developing social belonging.

Further, educators should focus on exposing their students to the numerous pathways within ML/AI. Particularly due to inequitable outcomes of biased ML/AI systems, we cannot afford to accept meritocratic excuses for the occupational gender segregation in this field. Most participants described ML roles as being deeply technical, meaning individuals who want to work in a less technical capacity may be less likely to persist. This hints at the "flipped hierarchy": within engineering, there is a tendency for women to end up in managerial roles, where less value is placed, rather than technical ones [50]. Thus, we might see fewer women pursuing highly technical ML roles. Understanding these role differences is important in uncovering barriers women face in ML/AI.

Additionally, as the high skills bar required for entry-level positions rewards personal projects and tinkering, it is likely harder for candidates motivated by the bigger picture implications of ML/AI technology to develop skills on their own. Yet, it is critical to retain socially-oriented individuals for an equitable future in ML/AI. Given that women, on average, value social benefit more in their work [18], [38], retaining socially-oriented individuals in general may also help increase diversity. Thus, educators should spend time describing areas that may be more enticing to candidates inspired by applications of ML/AI, such as in healthcare or business, as well as bringing in speakers working in ML/AI strategy, product, or business focused roles to broaden students' perception of work in ML/AI.

Lastly, our participants described that programming confidence, or a lack of, influences their persistence in ML/AI. Students described introductory courses as intimidating, especially if they did not have previous knowledge, and argued for the importance of giving everyone time to hone these skills. Thus, future work should examine the effect of introductory programming and ML/AI courses on students' persistence, and evaluate possible remedies like mindset or social belonging interventions, which have been effective in other contexts [48], [51]. Educators should consider offering multiple introductory programming courses that meet students where they are, creating a safe space for questions. Since side projects help students build programming confidence, educators can build these application-oriented assignments into course curricula.

For the research community, the insights from this early research cycle work highlight areas and recommendations for further study when understanding student experiences in ML/AI. In particular, the following recommendations are made to future ML/AI persistence research:
1. Questions about persistence in ML/AI should be qualified to understand which kinds of ML/AI roles a student sees themselves in. In our study, participants seem to have strikingly different ideas of what it means to persist in ML/AI, and future work should continue to capture this. In quantitative studies, ML/AI should be clearly defined, with examples of roles.
2. Questions about social belonging should assess the ability for students to develop meaningful *close* social connection, such as mentorship, rather than general social connection.
3. Persistence research in tech-heavy engineering fields should assess programming self-efficacy, which women report lower levels of [52]. The increasing demand for these skills in engineering may exacerbate the technical confidence gap, further perpetuating the gender gap in these fields.

We report on some initial ways persistence in ML/AI seems to be influenced by students' social belonging confidence. We motivate further work in this area to expose opportunities for institutions and workplaces to incorporate targeted support for underrepresented people in ML/AI, specifically through the following research questions:
- *How are social connections through close mentorship formed and how can these connections be fostered as a form of social belonging intervention in ML/AI?*
- *How do intersectional identities shape social belonging and persistence in ML/AI?*
- *How do programming or ML/AI courses inspire a fixed or growth mindset in students?*

Some of the reasons our participants described for choosing to persist in ML/AI differed from engineering persistence research more broadly. Because ML/AI is still rapidly evolving, some students are less concerned about masculine cultures, but more concerned about the ethical challenges within the field and consider this when choosing a career. We argue for the continued study of persistence by subfields and investigation into new fields as they emerge and evolve.

**Limitations**

This exploratory study resides in the early stages of the research cycle. Due to the small sample size, we are unable to draw sweeping, generalizable conclusions regarding how social belonging influences persistence in ML/AI. However, Lieberman argues that answering difficult questions requires moving from exploratory, descriptor studies through to experimental studies, and publishing early research cycle work incentivizes the discovery of new phenomena [53]. The insights presented here represent the start of important considerations for educators, and fruitful directions for future persistence research. Further, because participants were recruited on a voluntary basis, there may be self-selection bias in our sample. For example, the sample averaged extremely high in social benefit interest. Thus, the findings from this paper are skewed towards high social benefit interest individuals, and future work is needed to determine if these findings represent ML/AI students in general. Lastly, this study represents the experiences of students entering ML/AI from engineering. Differences in education, culture, and exposure may influence students' development of social belonging in the field. For example, CS students may have different experiences developing programming skills, thus experiencing fewer barriers when entering ML/AI. Future work should examine social belonging and persistence in ML/AI students with other backgrounds.

**Welcoming More Students Inside**

In search of an initial understanding of students' persistence in ML/AI, eight students were interviewed about their experiences, attitudes, and sense of social belonging. The most persistent participants in our sample discussed being independently motivated, highly interested in technical roles, and confident in their programming skills. These participants reported that they may have a harder time finding social connection, but receiving recognition for their work and a strong interest in the technological components help to make up for potential belonging uncertainty. Those in our sample who described themselves as less persistent shared struggling with programming, being more community oriented, being interested in the applications of ML/AI, and seeing ML/AI more as a useful skill than a career. These participants may require social connections to stay motivated. This early research suggests important directions for future work, including investigation into mentorship roles in ML/AI, the study of intersectional identities and persistence in ML/AI, and a targeted study of the impact of foundational programming courses on students' mindsets. As we continue to understand the reasons students choose to remain in or leave ML/AI, we can design educational spaces that encourage broad participation. Only when people of all identities have an equal voice in the field can we move towards an equitable future where ML/AI serves society at large.

# Appendix

**Appendix A: Summary of Interview Participant Responses to Independent Variables from Survey**

| Variable | Mean | Deviation from survey sample* [38] |
|---|---|---|
| **Self-assessment** | | |
| Non-technical self-assessment | 3.9 | +0.2 |
| **Confidence** | | |
| Career-fit Confidence | 2.5 | -<0.1 |
| Expertise Confidence | 3.2 | +0.3 |
| Social Belonging Confidence | 2.6 | -<0.1 |
| **Other Variables** | | |
| Social benefit interest | 4.8 | +0.4 |
| Toxicity of environment | 4.2 | -0.1 |
| Importance of a high salary relative to skills and experience | 4.1 | +<0.1 |
| Competitive Participation | 3.5 | +0.3 |
| Influence of popularity of ML/AI on course choice | 3.3 | -0.4 |

*Table A1: Summary of survey responses to key independent variables*
*\* All deviations from the mean were within the standard deviations of the larger study*

**Appendix B: Measures used in the related survey**

| Survey Questions | Variable measured |
|---|---|
| *Measured on a 5-point Likert scale from "Very Unlikely" to "Very Likely"* | |
| 1. What is the likelihood that you will take another ML/AI course in university? | Short-term Persistence |
| 2. What is the likelihood that you will be in an ML/AI role (academia or industry) in 5 years? | Long-term Persistence |
| *Measured on a 4-point Likert scale from "Not confident at all" to "Very Confident"* | |
| 3. ML/AI is the right profession for me. | Career-fit Confidence |
| 4. I can select the right role in ML/AI for me. | |
| 5. I can find a satisfying job in ML/AI. | |
| 6. I am committed to ML/AI, compared to my ML/AI classmates | |
| 7. I will develop useful skills through working with ML/AI. | Expertise Confidence |
| 8. I will advance to the next level of my career in ML/AI. | |
| 9. I have the ability to be successful in my career in ML/AI. | |
| 10. I will find community in the field of ML/AI. | Social belonging Confidence |
| 11. I will fit in with the professional culture in the field of ML/AI. | |
| 12. I will be able to relate to others in the ML/AI professional community. | |

*Measured on a 5-point rating scale from "Lowest 10%" to "Highest 10%"*

13. Rate your communication skills (e.g. writing and presenting) compared to an average person your age  *Self-assessment: Non-technical*
14. Rate your teamwork skills compared to an average person your age.
15. Rate your leadership abilities (eg. planning, delegating, and coordinating) compared to an average person your age.

*Measured on a 5-point Likert scale from "Strongly Disagree" to "Strongly Agree"*

16. It is important to me to do work that makes a helpful contribution to society; makes a difference.  *Social Benefit Interest*
17. It is important to me to do work that is consistent with my moral values.
18. It is important to me to work in an environment where workplace policies are administered with fairness and impartiality.

19. It is important to me that I earn a high salary (I.e., high relative to typical salaries for those with my skills and credentials) in my career.  *Earning Potential*
20. Now or in the recent past, I choose to participate in competitive events or activities (for instance: competitive athletics, judged performances, contests for funding or awards, entrepreneurial competitions, etc.)**  *Competitive Participation*
21. I was influenced to take this course because of the popularity of ML/AI as a topic of study.  *Interest in ML/AI*

22. I have experienced discrimination in some form in my ML/AI courses, in ways such as, but not limited to:  *Toxicity of Environment*
    - Direct (unequal treatment based on race, colour, sex, etc.)
    - Indirect or Systemic (driven by discriminatory policies or practices)
    - Harassment (unwelcome comments or actions)

    Which had the impact of excluding me, denying me benefits, or imposing a burden on me
23. I have noticed differences in the way I am spoken to in my ML/AI courses compared to my peers of a different identity.
24. I can identify instances in my ML/AI courses where negative stereotypes regarding my identity, academic standards for my identity, and/or expectation of ability of my identity, were reinforced.

**Appendix C: High-level codes and theme mapping**

| High-level code | Theme 1: Exposure and Initiation | Theme 2: Perceptions of careers in ML/AI | Theme 3: Social belonging and motivation | Theme 4: Importance of social connection through close mentorship | Theme 5: Perceptions of engrained and changing cultures within the discipline |
|---|---|---|---|---|---|
| Attitudes towards ML/AI (e.g., Belonging, Lack of confidence, Satisfying, Common ground, Self-learn) | | X | X | | |
| Community (e.g., Academic, AI, Conferences) | | | X | | |
| Courses (e.g., Mismatch with Industry, Hands-on, Previous knowledge, Research, Theoretical) | X | X | | | X |
| Definitions (e.g., Algorithm, AI, Efficiency, ML) | | X | | | |
| Ethics of AI (e.g., avoid ethical issues, Fairness, Human judge, Improve ethics) | | X | | | |
| Experience (e.g., Hands-on, Jobs) | | | X | X | |
| Interest (e.g., Career trajectory, Exposure to AI, Motivations, Industries) | X | X | X | X | X |
| People (e.g., Peer, Mentorship, Marginalized identity, Student culture, Approaching professors) | X | | | X | X |
| Resources (e.g., Online blog, Coursera, Faculty talks, Newsletters) | X | | | | |
| Skills (e.g., Programming, Skills for industry, Communication, | X | X | | | |
| Statistics, Technical, Writing) | | | | | |
| Technology (e.g., Computer Vision, Natural Language Processing, Prediction Model, Finance, Robotics) | | X | | | |
| Type of Work (e.g., Data Analysis, Data Science, Computer-focused, Working with people) | | X | X | | |
| Work Culture (e.g., Personality, Boundaries, Friendly, Lasting social connection, Male-dominated, Meritocracy, Open Communication) | | X | X | X | X |

Note the most commonly used low-level codes (by number of participants) are listed beside each high-level code. At least five are listed for high-level codes that have five or more low-level codes, more than five are listed in the case of ties.